\documentclass[preprint]{aastex} 

\newcommand{\com}[1]{}
\usepackage{amsmath,amssymb,amsfonts}
\usepackage{graphicx}
\usepackage[update,prepend]{epstopdf}
\usepackage{hyperref}
\usepackage{array}

\usepackage{cite}

\usepackage{natbib}
\begin{document}

\title{Observations of Giant Pulses from Pulsar PSR~B0950+08 using LWA1}
%
%
%

\author{
Jr-Wei Tsai\altaffilmark{1},
John H. Simonetti\altaffilmark{1},
Bernadine Akukwe\altaffilmark{2}, 
Brandon Bear\altaffilmark{1},
Sean E. Cutchin\altaffilmark{3},
Jayce Dowell\altaffilmark{4},
Jonathan D. Gough\altaffilmark{5},
Jonah Kanner\altaffilmark{6},
Namir E. Kassim\altaffilmark{7},
Frank K. Schinzel\altaffilmark{4},
Peter Shawhan\altaffilmark{8},
Gregory B. Taylor\altaffilmark{4}, 
Cregg C. Yancey\altaffilmark{8},
Leandro Quezada\altaffilmark{2},
Michael Kavic\altaffilmark{2}
}

\altaffiltext{1}{Department of Physics, Virginia Tech, Blacksburg, VA 24061, U.S.A}
\altaffiltext{2}{Department of Physics, Long Island University, Brooklyn, New York 11201, U.S.A.}
\altaffiltext{3}{NRC-NRL Postdoctoral Fellow, Washington, DC. 20375, USA}
\altaffiltext{4}{Department of Physics and Astronomy, University of New Mexico, Albuquerque NM, 87131.}
\altaffiltext{5}{Department of Chemistry and Biochemistry, Long Island University, Brooklyn, New York 11201, U.S.A.}
\altaffiltext{6}{LIGO-California Institute of Technology, Pasadena, California 91125}
\altaffiltext{7}{Naval Research Laboratory,  Washington, DC 20375, USA}
\altaffiltext{8}{Department of Physics, University of Maryland, College Park, MD 20742, USA}

\begin{abstract}
We report the detection of giant pulse emission from PSR~B0950+08 in 24 hours of observations made at 39.4~MHz, with a bandwidth of 16 MHz, using the first station of the Long Wavelength Array, LWA1.  We detected 119 giant pulses from PSR~B0950+08 (at its dispersion measure), which we define as having SNRs at least 10 times larger than for the mean pulse in our data set. These 119 pulses are 0.035\% of the total number of pulse periods in the 24 hours of observations.  The rate of giant pulses is about 5.0 per hour. The cumulative distribution of pulse strength $S$ is a steep power law, $N(>S)\propto S^{-4.7}$, but much less steep than would be expected if we were observing the tail of a Gaussian distribution of normal pulses. We detected no other transient pulses in a dispersion measure range from 1 to 90~pc~cm$^{-3}$, in the beam tracking PSR~B0950+08.  The giant pulses have a narrower temporal width than the mean pulse (17.8~ms, on average, vs.\ 30.5~ms). The pulse widths are consistent with a previously observed weak dependence on observing frequency, which may be indicative of a deviation from a Kolmogorov spectrum of electron density irregularities along the line of sight. The rate and strength of these giant pulses is less than has been observed at $\sim$100~MHz.  Additionally, the mean (normal) pulse flux density we observed is less than at $\sim$100~MHz.  These results suggest this pulsar is weaker and produces less frequent giant pulses at 39~MHz than at 100~MHz. 
\end{abstract}


\pagebreak 

\section{Introduction}

Since their discovery, pulsars have served as interesting astrophysical laboratories for the exploration of many phenomena.  Among their interesting features are giant pulses (GPs) and anonymously intense pulses (AIPs).  GPs or AIPs have been observed in only a handful of pulsars. The discovery of the Crab Pulsar was actually through the observations of its GPs \citep{1968Sci...162.1481S}.  There is some ambiguity in the definition of GPs and AIPs.  In a recent paper, \citet{2011A&A...525A..55K} define GPs as having flux densities which exceed the flux density of the mean pulse by at least a factor of 10, have a much narrower temporal width than the mean pulse, and have a power-law pulse intensity distribution. {\citet{2006ChJAS...6b..41K} found GPs from millisecond pulsars have a phase-alignment with X-ray emission. \citet{UlyanovZakharenko2012} discuss the difference between GPs and AIPs, with AIPs appearing at lower frequencies, and exhibiting narrrower-band emission than GPs.

{\citet{2012AJ....144..155S} define GPs as pulses with flux densities exceeding 10 times the mean pulse flux density from PSR~B0950+08. In this paper we present observations of pulses from PSR~B0950+08 with large intensity relative to the mean pulse, i.e., pulses with signal-to-noise ratio, SNR, at least 10 times that of the mean pulse, equivalent in practice to the GP definition adopted by \citet{2012AJ....144..155S}.  For simplicity, brevity, and comparison with previous work on PSR~B0950+08, we will refer to our large intensity pulses as GPs.

The search for radio transient pulses from astrophysical phenomena, including GPs, has benefited from additional instrumentation with the development of new  low-frequency arrays, such as the First Station of the Long Wavelength Array (LWA1) \citep{Ellingson:2012rc} and the Low-Frequency Array (LOFAR) \citep{2013A&A...556A...2V}.  These arrays possess the capability to observe astronomical sources at frequencies that have not yet been explored in great detail. Notably, LWA1 has been shown to be capable of detecting giant pulses emitted by the Crab Nebula pulsar between 20 and 80 MHz \citep{2013ApJ...768..136E}.  LOFAR has also been active in finding GPs at low frequencies;  \citet{2011A&A...530A..80S} report detections of bright single pulses in, e.g., PSR~B0950+08. Also a prior measurement of the polarization of average pulses from PSR~B0950+08 was conducted at 39 MHz \citep{1983AZh....60..554S}. An additional motivation for our current work is to prepare for using the LWA1 in searches for single dispersed pulses (fast transients) from a variety sources, including the possible radio pulse produced by a binary neutron star merger. Such mergers are the prime initial target of the LIGO-VIRGO collaboration \citep{2012PhRvD..85h2002A}; a coincident observation of a radio pulse from a merger could help in finding and identifying the gravitational wave signal from the event \citep{2010CQGra..27h4018P}.

GPs have been associated with PSR~B0950+08 in previous observations \citep{2004ApJ...610..948C, 2012AJ....144..155S, 2012ARep...56..430S}.  \citet{2012ARep...56..430S} has observed GPs for this pulsar at 112 MHz, during observing sessions of 3 minutes, over 22 days, finding pulses with strengths up to two orders of magnitude larger than the mean pulse strength.  \citet{2012AJ....144..155S} observed this pulsar at 103 MHz daily for half an hour over ten months.  They found that $\sim 1\%$ of the total pulses had a strength greater than 20 times the mean flux density, and one in ten thousand was greater than 100 times the mean flux density.  Their observations also suggested that the GP distribution is not uniform, with extremely active and silent days. \citet{2012AJ....144..155S} defined GPs in their observed data as pulses with flux densities at least 10 times as large at the mean pulse flux density, while \citet{2012ARep...56..430S} uses a factor of 5 as the threshold for GPs.  We will adopt 10 times the mean pulse strength as the threshold for GPs in our data.

Although giant pulses (GPs) have been widely detected and discussed in the literature \citep{1996ApJ...457L..81C,1538-4357-557-2-L93,1538-4357-590-2-L95, 2004A&A...413.1065N}, the mechanism for GP production remains unclear.   ~\citet{2004A&A...424..227P} suggests that GPs are created by induced scattering of pulsar emissions by particles of ultra-relativistic plasma in the magnetic field, possibly explaining the observations of \citet{2012ARep...56..430S}.  Other proposed mechanisms aim to explain the observed power-law statistics through wave collapse ~\citep{1996PhPl....3..192R}, though concerns have been raised about the potential of the wave collapse model \citep{2004ApJ...610..948C}.  \citet{2004ApJ...610..948C} argues for stochastic growth theory, though he admits it is not favored. It is still to be determined whether the power-law features observed in GPs are intrinsic or due to convolution effects. There is broad consensus in the literature that more observations at various frequencies are required to reduce the list of possible mechanisms.

In this paper we discuss 39.4 MHz observations we made of pulses from PSR~B0950+08 using LWA1. In Section 2 we discuss the LWA1 and our observations.  Section 3 details how we produced spectrograms from the data, and conducted a preliminary search for individual pulses across a large range of dispersion measures (DMs), finding statistically significant pulses only at the DM of PSR~B0950+08.  Then, in Section 4, we explain how we performed a detailed searched of our full data set for pulses at the DM of the pulsar, and measured the strength of each pulse; we specify pulse strength by SNR relative to the SNR of the mean pulse.  We also discuss the statistics of our observed pulses in this section.  In Section 5, we assign approximate flux densities to the mean pulse and GPs, and compare these results to those of previous studies.  Finally, we present our conclusions in Section 6.

\section{Observations}

LWA1 \citep{Ellingson:2012rc} is a new radio telescope operating in the frequency range 10--88~MHz, located in central New Mexico.  The telescope consists of 256 dual-polarized dipole antennas distributed over an area of about 110~m by 100~m, plus 5 outliers at distances of 200--500 m from the core of the array, for a total of 261 dual-polarized antennas.  The outputs of the dipoles are individually digitized and can be formed into beams (DRX beam-forming mode).  Four fully independent dual-polarization beams capable of pointing anywhere in the sky are available; each beam has two independent frequency tunings (selectable from the range 10--88~MHz) of up to about 17~MHz. The full-width at half-maximum (FWHM) beamwidth for zenith pointing is approximately $4.3^\circ$ at 74~MHz and depends on observing frequency as $\nu^{-1.5}$.   The system temperature is dominated by Galactic emission and so the beam sensitivity of the instrument is dependent on the LST of the observation and the direction of the beam.

We used LWA1 to observe PSR~B0950+08 for 30 hours, in 6-hour blocks, in 5 days during late March and early April 2012.  Observations were started about 45 minutes before transit of PSR~B0950+08 in each observing session.  We tracked the pulsar with 2 beams.  Each beam had 2 frequency tunings, yielding observations centered at 19.8~MHz, 39.4~MHz, 59.0~MHz, and 73.7~MHz, each with a bandwidth of 19.6 MHz (the used bandwidth is 16~MHz after data reduction).  The observations were done during the commissioning period for LWA1.  In the end, only 24 hours of observations (4 six-hour observing blocks) for one beam, centered on 39.4~MHz, were analyzed for this paper.  The observing sessions analyzed here are for the observing dates March 18, 24, 25, and April 1.  
These were the cleanest data, and, by using them, we avoided some pointing-error issues which influenced the results at higher frequencies.  These observations were made before Initial Operating Capability was reached on April 24, 2012; thus the current instrument now has better sensitivity and the pointing issues have been fixed.

\section{Data Reduction \& Preliminary Pulse Search}

Using routines from the LWA Software Library (LSL) \citep{Dowell:2012rt}, we performed a 4096-channel Fast Fourier Transforms (FFTs) on each 0.209 ms of raw data, dividing the 19.6 MHz observing bandwidth into channels of 4.785 kHz.  RFI mitigation was performed on the data set, using the following procedure.  First, we obtained the average spectrum for each 2.09-s interval (each set of 10,000 consecutive spectra).  Next, we fit a 16th order polynomial to the 2.09-second average spectrum, and divided that average spectrum by the polynomial. The 16th order polynomial is the lowest order that fit the approximately 16 ripples in the bandpass, without undue suppression of narrow-band RFI. Finally, any frequency bin in the 2.09-s average spectrum that was greater than 3$\sigma$ above the mean was masked as RFI contaminated in all the corresponding 0.209-ms spectra.

The shape of the observed bandpass is not constant in time, and this variation must be removed to allow for an effective search for transient pulses.  The variation in the spectrum is dominated by the diurnal variation of the Galactic background.  Once the RFI-contaminated frequency bins were identified and masked, we determined and removed the varying shape of the bandpass using the following procedure.  First, for each 2.09 s of data, we computed a median spectrum for the 10,000 RFI-masked spectra.  Then, we used 150 such spectra to compute the median spectrum for approximately 5 minutes of observations. The 5 minute duration was chosen so the diurnal variation of the Galactic background was effectively smoothed out.  To further smooth the 5-minute spectrum, we performed a moving boxcar average across the 5-minute median-spectrum; the boxcar length used was 101 frequency channels.  Finally, we divided each 0.209-ms spectrum by the boxcar-smoothed spectrum corresponding to its epoch.  To prepare for further analysis we removed the first 360 channels and the last 395 channels from each spectrum, removing any end effects, and leaving a final bandwidth of 16.0~MHz.  The final spectra were arranged into spectrograms of frequency (vertical axis) and time (horizontal axis).
\citet{2013ApJ...768..136E} were able to visually identify Crab GPs in spectrograms. Figure.~\ref{signalintimeseriesandspectrum} shows a portion of a full spectrogram (only 2.4~MHz, not the full 16~MHz, and only a few seconds of data) displaying one of our stronger GPs, along with the corresponding time series obtained by dedispering the displayed data at the DM of PSR~B0950+08.

\citet{2003ApJ...596.1142C} describe in detail a technique suitable for searching for individual pulses of various origins, including pulsar giant pulses, in time-frequency data such as ours.  As a first look, we used this method on our data, taken in chunks of 5-minute duration. In essence, the technique consists of constructing dedispersed time series for a range of candidate DMs and smoothing each individual time series with effectively larger and larger averaging-boxcars to search for pulses of temporal width matched to the smoothing time --- which yields the best signal-to-noise ratio for a candidate pulse.  Pulses of strengths and numbers larger than expected for the (assumed) Gaussian noise in our data are pulses of possible astrophysical origin.  (Another possibility is that they are RFI or other transient non-astrophysical events, but candidate pulses of the correct DM for PSR~B0950+08 are more likely to be our sought-after pulsar pulses.)  We performed incoherent dedispersion (summing intensities) in our spectrograms. We searched through time series for 15,030 candidate DMs in this manner, ranging from 1.0 to 89.9 pc cm$^{-3}$. The DM spacing is given by 
\begin{equation}
\delta \mbox{DM} = \mbox{DM} \frac{\Delta\nu}{B},
\label{eqn:DM spacing}
\end{equation}
where $B$ and $\Delta\nu$ are the bandwidth and channel-width respectively. Thus the temporal smearing due to DM spacing is equal to temporal smearing across a frequency channel, which can not be removed. The time series was smoothed in steps by averaging a moving boxcar of width equal to 2 time samples, and then removing one of the resulting time samples. 
Repeated smoothing and decimating in this manner efficiently produces a set of time series of increasing smoothness. At each smoothing and decimation step the resulting time series ass searched for pulses.  We performed 15 such steps for each dedispersed time series. Thus, the final time sample duration in the last-smoothed time series was $2^{15}\times0.2089{\rm s} = 6.85$s.

A representative 5-minute result of this DM-space search is shown in Figure~\ref{GpsEventRate}.  In the entire search we found transient events in the resulting time series with SNRs $\gtrsim$ 6.5 were for a DM of 2.97~pc~cm$^{-3}$, the DM for PSR~B0950+08.  No such strong pulses were found at other DMs.  Furthermore, as indicated in the figure (and as explained by Cordes and McLaughlin), the expected number of transient events due to Gaussian noise matches well our numbers of events at SNR$<$6.5, but events of SNR$>$6.5 are more numerous than expected from Gaussian noise alone, and appear at the DM of the pulsar.  Thus we are confident that by focusing on transient events that have a SNR$>$6.5, determined through this procedure, we are selecting pulses produced by PSR~B0950+08.  

We note here that the SNR determined for a pulse by the Cordes-McLaughlin procedure is computed in the time series smoothed to the temporal width of the pulse. This is a precise means of quantifying the SNR of a temporally-isolated \textit{single, dispersed pulse}, and, as such, is perhaps reasonable for describing anomalously intense pulses or giant pulses, which tend to be isolated.  But, it should be noted that quoted pulsar flux densities are averages in time, including both energy received during pulses, \textit{and} between pulses, i.e., effectively zero.  Thus, when we measure the SNR of our large pulses and compare them to the SNR of the mean pulse, we will adopt the more conventional time-average throughout a pulse period.  

Finally, to further explore the parameters necessary to conduct an effective and efficient search for large individual pulses from PSR~B0950+08, we searched one 5-minute sequence of bandpass-corrected 0.209~ms spectra using an incoherent dedispersion routine with trial DMs ranging from 2.965 to  2.975~pc~cm$^{-3}$.  The transient time-sample events with highest SNR in the resulting time series were obtained for a DM = 2.970 pc cm$^{-3}$.  Thus, we narrowed our subsequent search for GPs within the entire data set to this value of DM.

\section{Pulses from PSR~B0950+08}

\subsection{The Mean Pulse}


The mean pulse profile is a sum of the ``folded" time series for the entire 6-hours. The folding occurred at the pulse period (253.077573~ms) and was optimized by using PRESTO \citep{RansomThesis} after converting the data into the PSRFITS format. This pulse profile is also smoothed such that there are 64 time bins across the pulse period (each time sample is about 4~ms).  The normal pulses would be undetectable individually (in the sense discussed in the previous section), but the mean pulse is detectable due to the large number of folds possible in the data set.  This mean profile is in agreement with the normal pulse profile for PSR~B0950+08 in pulse period, pulse width, and pulse shape \citep{2011A&A...530A..80S}.  The main purpose in running PRESTO independently on our data was to verify the data reduction software used for this work. For subsequent analysis, we used programs we wrote, for increased flexibility in analysis.  Our mean pulse profiles are consistent with the PRESTO result, as discussed below. In the PRESTO mean pulse profile, and in our mean profiles, it is apparent that there are two components.  These components are closely separated in time.  In this paper we characterize the strength and width of the mean, and giant pulses, by fitting a single Gaussian to a pulse. This is justified in part by the lack of scatter broadening observed in the giant pulses with a single peak. This especially simplifies the analysis of \textit{single} giant pulses, where the noise in the baseline is stronger relative to the peak of a pulse, than in the case of the mean pulses shown in Figure~\ref{PRESTOprofile} and Figure~\ref{meanpulses}.

The PRESTO output indicated that the strength of the pulsar appeared to systematically decrease throughout the 6-hour observing session. We investigated this behavior more carefully by writing a program to fold and sum the data in 5-minute chunks throughout the observing sessions, without smoothing the time series, using the pulse period optimized by using PRESTO.  Figure~\ref{meanpulsevstime} shows the resulting average SNRs during a pulse period as a function of time, for three of the 6-hour observing sessions.  The systematic behavior in each observing session is the same, peaking about 40-45 minutes after the start of a session, corresponding to the transit of PSR~B0950+08.  We interpret this behavior as indicating a decrease in beam sensitivity as a function of increasing zenith angle.  Note that by watching the behavior of the SNR, we automatically include both the decreased effective collecting area of the array with increasing zenith angle, and the increased system noise with increasing zenith angle (see Section 5).  We fitted a third-order polynomial through these data, and normalized the SNR and polynomial values to unity at the maximum of the polynomial; we used this normalized curve to systematically correct all SNR measurements determined by subsequent analysis (multiplying pulse strengths by the appropriate factor which would make the SNR in the figure equal to the typical SNR at transit).  

The mean pulse profiles determined from folding our data during the first 90 minutes of each of the four observing sessions, are shown in Figure~\ref{meanpulses}.  Note the fairly consistent shape and strength of these mean pulse profiles; the consistent mean-pulse strength is apparent, by comparing the height of each pulse to the rms in each panel. A fitted Gaussian shape is used to measure the mean pulse strength and width for each panel. The mean pulse has a FWHM of about 30.5~ms.  The SNR of each displayed mean pulse is clearly quite high.  We took the average SNR throughout the full pulse period for each of these displayed pulses (signal $=$ sum of the fitted Gaussian values, rms $=$ rms in the baseline), and averaged the four results (which were within 10 percent of each other), then reduced the average by $\sqrt{N_{\rm folds}}$, where $N_{\rm folds}$ is the number of folds necessary to produce the profiles shown, obtaining the SNR of the mean pulse for our observations, SNR$_{\rm mean\ pulse}$.  This is the SNR that can be assigned to a \textit{single, typical pulse}, and can be used to characterize the SNR of any individual large pulse we detected, through the ratio SNR/SNR$_{\rm mean\ pulse}$.  Even in the absence of accurate flux density values for the mean pulse and individual large pulses, this ratio is a reasonably precise characterization of an individual large pulse's strength.

\subsection{Individual Large Pulses}

Four six-hours dedispersed time series (DM=2.9702 pc cm$^{-3}$), with 0.209~ms time samples, were constructed by averaging the times series for the two orthogonal polarizations.  Reading data, and a few other organizational steps, was carried out using LSL routines \citep{Dowell:2012rt}. These time series were searched for individual large pulses using the following procedure.  First, we found each individual time sample with a signal-to-noise $>$3.5.  Then, we fitted a Gaussian to the 1000 time samples centered on the $>$3.5$\sigma$ sample, and obtained the FWHM of the resulting Gaussian. We isolated the subset of time samples of total duration equal to 2000 FWHM as the baseline, which centered on the candidate pulse. Next, we decimated this time series by computing average values in intervals of length equal to the FWHM, making sure one interval was centered on the $>$3.5$\sigma$ peak.  Then, we computed the SNR of the potential GP in the decimated time series.  If this SNR was $>$6.5, we recorded this event as a pulse.  Note here the SNR for the event is taken to be the SNR in the decimated time series, in the sense used in the Cordes-McLaughlin technique, ensuring we are capturing pulses that are not part of the Gaussian noise in our time series data.  As a further check on the reality of each pulse detected in this manner, we only kept candidate pulses with a FWHM that fell in a nominal range of 4.178~ms to 41.78~ms (20 to 200 of the 0.209-ms time samples); this pulse width range includes the width of the best candidate pulse in the initial 5 minutes of data we searched, and the duration of the double-peaked pulses we found in those data.  Finally, we avoided double counting GPs by keeping only candidates that were at least 40 ms away from any other candidate pulse.  An example pulse is shown in Figure~\ref{gpidentifygaussianfit}.  Some dual-peak pulses are shown in Figure~\ref{Gps2peaks}.

\subsection{Results}

Using the methods discussed above, we detected 398 individual pulses in the 24 hours of observed data; these are pulses that are strong enough to not be considered part of the Gaussian noise in the time series data. Using the definition of \citet{2012AJ....144..155S} for giant pulses (SNR/SNR$_{\rm mean\ pulse} > 10$), the number of such GPs is 119, or about $119/(24\times3600/0.253)\approx0.035$\% of the total number of pulses that occur in the 24 hours. The maximum value of SNR/SNR$_{\rm mean\ pulse}$ for an individual pulse in our data set is 28.2, corresponding to a pulse of temporal width 19.6~ms (FWHM), considerably narrower than the pulse width of the mean pulse (30.5~ms). 

Nearly all the GPs have a double-peaked profile.  It may be that all are double-peaked, but one peak may be hard to discern given the noise in the single-pulse profile.  To measure GP pulse strength and width, we fit a single Gaussian to the profile, as we did for the mean pulse, which simplifies the analysis and allows for a valid comparison of the GPs with the mean pulse.  Figure~\ref{Gps2peaks} shows some GP profiles displaying two peaks.



Figure \ref{pulsenumberhistogram} displays a histogram of the detected pulses, in terms of SNR/SNR$_{\rm mean\ pulse}$.  GPs are those with SNR/SNR$_{\rm mean\ pulse} > 10$. The number of GPs decreases sharply with increasing strength.  Figure~\ref{widthhistogram} shows a histogram of the GP widths.  The average GP width (FWHM) is 17.8~ms, about half the width of the mean pulse.

Figure~\ref{loglogGpsEventRate} shows the cumulative distribution of pulse strength, $N(>S)$, where the strength $S$ is the relative SNR, i.e., $S=$~SNR/SNR$_{\rm mean\ pulse}$. The figure shows a power-law fit to the GPs only $S>10$) which yields $N(>S) \propto S^{-4.7}$.  It is apparent that the GPs we observed have a different distribution than pulses with relative SNR$<10$.  The distribution for the GPs is not as steep as would occur for pulses in the tail of a Gaussian distribution of normal pulses.  For a Gaussian probability distribution, the fraction $f(>z)$ of pulses with strength greater than $z\sigma$ from the mean is given by 
\begin{equation}
f(>z) = \frac{1}{2}\ {\rm erfc}\left(\frac{z}{\sqrt{2}}\right).
\end{equation}
To obtain $f(>z)=0.00035$, requires $z=3.39$, and $d\log f/d\log z \approx -12.4$ at $z=3.39$, much steeper than our value of $-4.7$.  Thus, the observed $N(>S)$ power-law implies the GPs are not in the tail of a Gaussian distribution of normal pulses.  

Interestingly, the $-4.7$ power-law is steeper than found by \citet{2012AJ....144..155S} ($-2.2$) or \citet{2012ARep...56..430S} ($-1.84$ at the steepest), at frequencies just over 100~MHz.  Their observations also show PSR~B0950+08 to be stronger (larger mean flux density), to produce stronger GPs (as discussed in the next section), and to produce GPs at a higher rate than for our observations at 39.4~MHz (\citet{2012AJ....144..155S} report $\sim$1\% of the pulses in their observing period were GPs).  All these results, taken as a whole, may indicate PSR~B0950+08 is a weaker producer of GPs at 39.4~MHz, than at $\sim$100~MHz.  \citet{2012AJ....144..155S} found there was a large day-to-day variation in the rate of GPs they observed with more than 99\% of their GPs occurring during $\sim$25\% of their observing sessions.  By contrast, the number of GPs we observed over the four separate 6-hour observing sessions were 11, 41, 31, and 36, thus the same (to within root $N$) for three of the four observing sessions.  Apparently, PSR~B0950+08 is also less variable in its GP output rate at 39.4~MHz, according to our observations.

\section{Flux Densities}

Our observations did not include any drift scans of strong flux density, so we obtain rough flux densities for the mean pulse and our individual larger pulses using an estimated system equivalent flux density (SEFD).  The SEFD is the flux density a source in the beam needs to produce a SNR of unity, for an observation of 1~Hz bandwidth and integration time of 1~second.  At low frequencies, Galactic noise is the dominant contribution to system noise.  Ellingson established a rough model for estimating the SEFD, which takes account of the combined effects of all sources of noise \citep{Ellingsonsen}. Ellingson uses a spatially uniform sky brightness temperature $T_b$ in his model, dependent on observing frequency $\nu$, where
\begin{equation}
T_b = 9751 \mbox{K} \left(\frac{\nu}{38\mbox{MHz}}\right)^{-2.55} 
\label{eqn:brightness temperature}
\end{equation}
and ignores the ground temperature contribution as negligible. The receiver noise is about 250~K, but has little influence on the SEFD. This model when applied to LWA1 shows that the correlation of Galactic noise between antennas significantly desensitizes the array for beam pointings that are not close to the zenith.  It is also shown that considerable improvement is possible using beam-forming coefficients that are designed to optimize signal-to-noise ratio under these conditions. \citet{Ellingson:2012rc} checked this model with observations of strong flux density calibrators, finding the results roughly correct.  Based on the model of Ellingson and his drift scan results, and given our observations of PSR~B0950+08 at transit are for a zenith angle of about 26$^\circ$, we estimate that an appropriate SEFD to use for our observations at transit is 15,000~Jy, with an uncertainty of roughly 50\%.  The SNR of pulses away from the moment of transit are corrected by a factor which compensates for decreasing effective collecting area and increasing SEFD, with increasing zenith angle.

Thus, the flux density we assign to a pulse, as averaged across the entire pulse period, is
\begin{equation}
S_\nu = \frac{\rm SEFD}{\sqrt{2B\Delta t}}\ \frac{1}{N_{\rm bins}} \sum_{i=1}^{N_{\rm bins}} \frac{I_i}{\rm rms} = \frac{\rm SEFD}{\sqrt{2B\Delta t}}\ {\rm SNR}
\label{eqn:flux density}
\end{equation}
where SEFD = 15,000~Jy, $B=16\times10^6$~Hz is the bandwidth, $\Delta t=0.2089\times10^{-3}$~s is the duration of a time sample, $N_{\rm bins} = 1211$ is the number of time samples (bins) in a pulse period of 253~ms, the sum is over the full pulse period, the $I_i$ are the intensity values (arbitrary units) in the Gaussian pulse profile fitted to a pulse (a baseline average was already subtracted from the data), rms is measured in the baseline, and the SNR is the average signal-to-noise ratio during the pulse period.  Taking the bandwidth as a full 16~MHz is appropriate, despite the masking of RFI in the spectrograms, since less than 1\% of the data were lost to RFI.

In determining pulse flux densities we made no corrections for the LWA1 pointing errors known to be present at the time of the observations.  \citet{LWAmemo194} indicates that telescope operators were making manual adjustments to pointing in late March 2012, and then starting on April 4, 2012, corrections were made automatically before running any observing file.  These corrections were $-$7 minutes in RA and +1$^\circ$ in declination; at the declination of PSR~B0950+08 this amounts to a total correction of about 2$^\circ$.  Assuming those pointing corrections may not have been applied during our observations (and should have been), we can estimate the resulting decrease in flux density that would have resulted.  Assuming a beam FWHM of 4.3$^\circ\times(39.4/74)^{-1.5}$ \citep{Ellingson:2012rc}, or $\approx 11.1^\circ$, and a Gaussian beam shape, the measured flux density would be 98\% of the true flux density, a negligible difference, especially given our 50\% uncertainties in flux density.

The mean pulse for our observations has a flux density of $1.5\pm0.75$~Jy.  Since our ``threshold'' for declaring the detection of a giant pulse is set at SNR/SNR$_{\rm mean\ pulse} > 10$, the corresponding flux density is $15\pm7.5$~Jy.  Our pulse with largest SNR/SNR$_{\rm mean\ pulse} = 28.2$ has a flux density of $42.3\pm21$~Jy.  

Figure \ref{SpectralIndex} shows flux densities for normal pulses from PSR~B0950+08 at frequencies of 408~MHz and above, along with observations of the mean and giant pulses by \citet{2012AJ....144..155S} at 103~MHz  and \citet{2012ARep...56..430S} at 112~MHz.  In addition, we have added data points for the mean and giant pulses we observed.  The error bar on our mean pulse flux density indicates our 50\% uncertainty.  The range of our GPs are indicated by a solid red vertical line, with upper and lower error bars indicated.  Apparently our mean pulse and giant pulses are weaker than the those found by \citet{2012AJ....144..155S} at 100~MHz, and \citet{2012ARep...56..430S} at 112~MHz. This is consistent with the work of \citet{1994A&A...285..201M} where it was found that the mean flux density at 39.4~MHz appears to deviate more from the power law fit to higher frequencies than occurs at $\sim$100~MHz. As we noted in Section 4.3, the cumulative distribution of pulse strength, $N(>S)$, has a steeper power-law for the GPs than found by \citet{2012AJ....144..155S}, and \citet{2012ARep...56..430S}.  All these results perhaps indicate that PSR~B0950+08 is weaker, and produces less frequent and less intense giant pulses at 39~MHz than at 100~MHz.

\section{Scatter Broadening}

Spatial variations in the interstellar free-electron number density are responsible for the scattering and scintillation of radio signals propagating through the interstellar medium.  Pulsar observations are particularly useful for measuring the effects of interstellar scattering, and therefore characterizing the interstellar electron-density irregularities.

\com{Analysis of the effect of scattering typically assumes a homogeneous isotropic turbulence screen and fluctuation scales within a valid range so as to avoid chaotic behavior \citet{1977ARA&A..15..479R}. Then the power spectrum of the scattered wave can be expressed as
\begin{equation}
P_{ne}=\frac{C^2_{ne}}{(q+k^2_o)}\rm exp\left (\frac{q^2}{4k^2_i}\right ),
\end{equation}
where $q$ is the wavenumber which is bound by $k_i$ and $k_o$ the inner and outer turbulence scales and $C_{ne}$ is the fluctuation strength for a given line of sight (LOS). If the wavenumber is indeed in the range $k_{o}\ll q\ll k_{i}$, the equation reduces to a simple power law,
\begin{equation}
P_{ne}=C^2_{ne}{q^{-\beta}}.
\end{equation}
For the Kolmogorov spectrum we expect $\beta = 11/3$ and the spectral index of the scattered broadened pulse width becomes $\alpha=2\beta/(\beta-2)$= 4.4 (e.g. \citet{1986MNRAS.220...19R}),
\begin{equation}
\Delta t_{scattering}\propto \Delta \nu ^{-\alpha }.
\end{equation}
However, there are observations of pulsars over a wide range of DMs which exhibit a departure from Kolmogorov. For example a small group of pulsars with a high DM range (582-1074 pc cm$^{-3}$) were observed to have an average index of 3.44$\pm$0.13 \citep{2001ApJ...562L.157L}), a larger group of low galactic-latitude pulsars were observed to have an average index of 3.9$\pm$0.2 \citep{2004ApJ...605..759B} as were the unknown sources of FRBs \citep{2013MNRAS.436L...5L,2012MNRAS.425L..71K}. The PSR~B0950+08 was found to exhibits an exceptionally small index as 0.55~\citep{1976ApJ...209..895B} below 300MHz. }

Analysis of interstellar scattering often uses a thin screen between the source and observer, containing electron number-density fluctuations with a power-law spatial power spectrum \citet{1977ARA&A..15..479R}.  For wavenumbers $q$ between the outer and inner scales of the irregularities, $q_o$ and $q_i$, i.e., $q_o \ll q \ll q_i$, the power-spectrum for the electron-density irregularities is often expressed as a power-law,
\begin{equation}
P_{n_e}=C^2_{n}{q^{-\beta}},
\end{equation}
where $C_{n}$ is the fluctuation strength for a given line of sight (LOS).
For a Kolmogorov spectrum $\beta = 11/3$ and the scattered-broadened pulsar pulse-width depends on observing frequency as (e.g., \citet{1986MNRAS.220...19R}),
\begin{equation}
\Delta t_{\rm scattering} \propto \nu^{-\alpha} \propto \nu^{-2\beta/(\beta-2)} \propto \nu^{-4.4}.
\end{equation}
A Kolmogorov spectrum is often assumed, and is consistent with many pulsar observations.
However, there are observations of pulsars over a wide range of DMs which exhibit a departure from Kolmogorov. For example, a small group of pulsars with a high DM range (582-1074 pc cm$^{-3}$) were observed to have an average $\alpha = 3.44 \pm 0.13$ \citep{2001ApJ...562L.157L}), a larger group of low galactic-latitude pulsars were observed to have an average $\alpha = 3.9 \pm 0.2$ \citep{2004ApJ...605..759B}.
The PSR~B0950+08 was found to exhibit an exceptionally small $\alpha= 0.55$~\citep{1976ApJ...209..895B}, below 300MHz. 

\com{
\citet{2004ApJ...605..759B} and \citet{2013MNRAS.434...69L} explored several plausible explanations for the departure from the $\alpha =4.4$.  
In order for $\alpha=4.4$ all four of the following conditions must be fulfilled. (i) The electron density spectrum is of the Kolmogorov form. (ii) Only a thin screen or a uniformly thick bulk is in the LOS. (iii) The wave number falls in the range between the inner and outer scales. (iv) The turbulence is isotropic and homogeneous. Thus there are a variety of reasons which could explain the departure from $\alpha=4.4$ and there exists observational evidence for pulsars which deviate from it under different circumstances. For example the truncation of electron distribution has been observed to cause a deviation \citep{2001ApJ...549..997C} as well as the wavenumber being smaller than inner turbulence scale \citep{1986MNRAS.220...19R} which causes a flatter spectral index for temporal pulse width dependence. Pulsars and FRBs which deviate from Kolmogorov tend to show a flatter scattering spectrum which makes low frequency observations more practical.}

\citet{2004ApJ...605..759B} and \citet{2013MNRAS.434...69L} explored several plausible explanations for the departure from $\alpha =4.4$. In order for $\alpha=4.4$ all four of the following conditions must be fulfilled. (i) The electron density spectrum is of the Kolmogorov form. (ii) Only a thin screen or a uniformly thick bulk lies in the LOS. (iii) The wavenumbers sampled by the observations fall in the range between the inner and outer scales. (iv) The turbulence is isotropic and homogeneous. Thus there are a variety of deviations from these conditions which could explain the departure from $\alpha=4.4$.  Some pulsar observations which deviate from Kolmogorov results have been attributed to deviations from the assumptions summarized above. For example, a truncation of the electron-density power spectrum has been proposed as a cause \citep{2001ApJ...549..997C}. It has also been proposed that observations on a spatial scale smaller than the inner turbulence scale can cause a deviation from Kolmogorov resulting in an $\alpha$ less than 4.4 \citep{1986MNRAS.220...19R}.  

\com{We folded the 4$\times$6 hours of observations of the mean pulse using the period optimized by PRESTO, as shown in Figure \ref{meanpulses}. We then fit the folded mean pulse profile by assuming a Gaussian function which yielded an average FWHM pulse width as 30.5$\pm$5.5 ms from the 4 observations, centered at 39.4MHz with the bandwidth of 16MHz. A temporal scatter broadening spectral index of 0.43$\pm$0.078 was found when our observations were placed in the context of other observations of this pulsar at frequencies from 25MHz to 410MHz as shown in Figure \ref{pulsewidthindex}. We also find a spectral index of 0.08$\pm$0.024 when fitting only the higher frequency observations from 410MHz to 10.7GHz. Our results which indicate a very small scattering index are consistent with other observations of the effect of scattering on pulses from PSR~B0950+08 . The most plausible reasons for the flattened spectrum of PSR~B0950+08 is the peculiar distribution of the electron density along the LOS. This was discussed in \citep{2003astro.ph..1598C}. They argued that PSR~B0950+08 has such a low scattering measure due to the path length being predominantly thorough the Local Hot Bubble (LHB) and the Local Super Bubble (LSB), which have small electron densities and small fluctuation parameters. Thus the plausible reasons for the flattened scattering of PSR~B0950+08 consistent with our observations should be the peculiar LOS which is through a region that either has inhomogeneous scattering screen, anisotropic turbulence or uncommon wavenumber scales due to the small electron densities.}

We folded the 4$\times$6 hours of observations of the mean pulse, as shown in Figure \ref{meanpulses}. We then fit the folded mean pulse profile by assuming a Gaussian function which yielded an average FWHM pulse width as 30.5$\pm$5.5 ms from the 4 observations, centered at 39.4MHz with the bandwidth of 16MHz. We find a temporal scatter-broadening spectral index of $\alpha = 0.43\pm0.078$ when our observations are placed in the context of other observations of this pulsar at frequencies from 25MHz to 410MHz, as shown in Figure \ref{pulsewidthindex}. Our results are consistent with other observations of the effect of scattering on pulses from PSR~B0950+08. Perhaps the most plausible reasons for the small $\alpha$ for PSR~B0950+08 is the peculiar distribution of the electron density along the LOS. This was discussed in \citep{2003astro.ph..1598C}. They argued that PSR~B0950+08 has such a low scattering measure due to the path length being predominantly thorough the Local Hot Bubble (LHB) and the Local Super Bubble (LSB), which have small electron densities and small fluctuation parameters. Perhaps the explanation for the low value of $\alpha$ for PSR~B0950+08 would be the peculiar LOS which is through a region that has an inhomogeneous scattering screen, anisotropic turbulence or an uncommon wavenumber spectrum.  Further observations of PSR~B0950+08 could help illuminate the circumstances that cause the observed deviation from the Kolmogorov result.

\section{Conclusions}

We observed PSR B0950$+$08 for 24 hours at 39.4~MHz with a bandwidth of 16 MHz, using the LWA1, determined a mean pulse profile, and detected 398 pulses strong enough to not be part of the tail of Gaussian noise in our data.  Of these pulses, 119 have a SNR greater that 10 times the SNR of the mean pulse.  \citet{2012AJ....144..155S} considered such strong pulses ``giant pulses'' in their observations of B09050+08 at 103~MHz. For a large number of giant pulses, we observed a double-peak structure, similar to that observed in the normal pulses for this pulsar. The mean pulse profile (found by folding) has a FWHM of about 30.5~ms.  The giant pulses are almost all narrower, having a FWHM, on average, of 17.8~ms.  These 119 giant pulses are about 0.0035\% of the total number of pulses that occur in the 24 hours. The rate of the GPs we observed is about 5.0 per hour. The cumulative distribution of giant pulse strength, $N(>S)$, follows a power law with power $-4.7$.  We assigned a rough flux density of $1.5\pm0.75$~Jy to the mean pulse, and our largest giant pulse is 28.2 times stronger than the mean pulse.  No other transient pulses were observed in a DM range from 1 to 90~pc~cm$^{-3}$.  Given that all transients were observed at the known DM of the pulsar there is a high degree of certainty that all observed transients were associated with the pulsar itself. The absence of transients allows for a limit to be set on the agnostic transient rate in this range of DMs, during our 24 hour beam on PSR B0950+08.

Our mean pulse flux density and giant pulse flux densities are weaker than observed by \citet{2012AJ....144..155S} at 103~MHz, and \citet{2012ARep...56..430S} at 112~MHz.  The GPs observed by \citet{2012AJ....144..155S} were $\sim1$\% of the pulses that occurred over their total observing time, a rate much higher than ours.  In combination with the steeper power-law cumulative distribution of giant pulse strengths we found for GPs at 39.4~MHz, we conclude we may be seeing a weaker and less active pulsar at 39.4 MHz than was observed at $\sim$100~MHz.  Further observations of PSR~B0950+08 at low frequencies would be useful. Observations of the Crab pulsar demonstrated that the strongest GPs tend to have a shorter duration \citet{2007A&A...470.1003P}. We did not see this trend in GPs from PSR B0950+08 in the frequency range observed.

Additional work of interest would include sensitive enough observations of PSR~B0950+08, and other pulsars which exhibit pulses of large strength, to simultaneously explore the detailed statistical distributions of normal pulses and giant pulses.  Such studies could put the observational identification of giant pulses on a firmer footing, and advance the theoretical modeling of giant pulses.

These observations demonstrate the usefulness of LWA1 in searches for transient pulses.  In particular, agnostic searches would be valuable, in addition to targeted work such as reported here. Wide-area LWA1 searches could also be leveraged to enable more sensitive LIGO-VIRGO searches for gravitational-wave events, triggered by the detection of radio transients. The benefits of such collaborative work will be described in another publication.

\acknowledgments

We would like to acknowledge insightful discussions with S.W. Ellingson, T.J.W. Lazio and P. S. Ray. Construction of the LWA has been supported by the Office of Naval Research under Contract N00014-07-C-0147.  Support for operations and continuing development of the LWA1 is provided by the National Science Foundation under grant AST-1139974 and AST-1139963 of the University Radio Observatory program.  Part of this research was performed while S.E. Cutchin held a NRC research appointment at NRL.  Basic research in radio astronomy at NRL is supported by 6.1 base funding.

{\it Facility:} \facility{LWA}

\bibliography{referencesp}

\begin{figure}
\begin{center}
\includegraphics[width=0.75\textwidth]{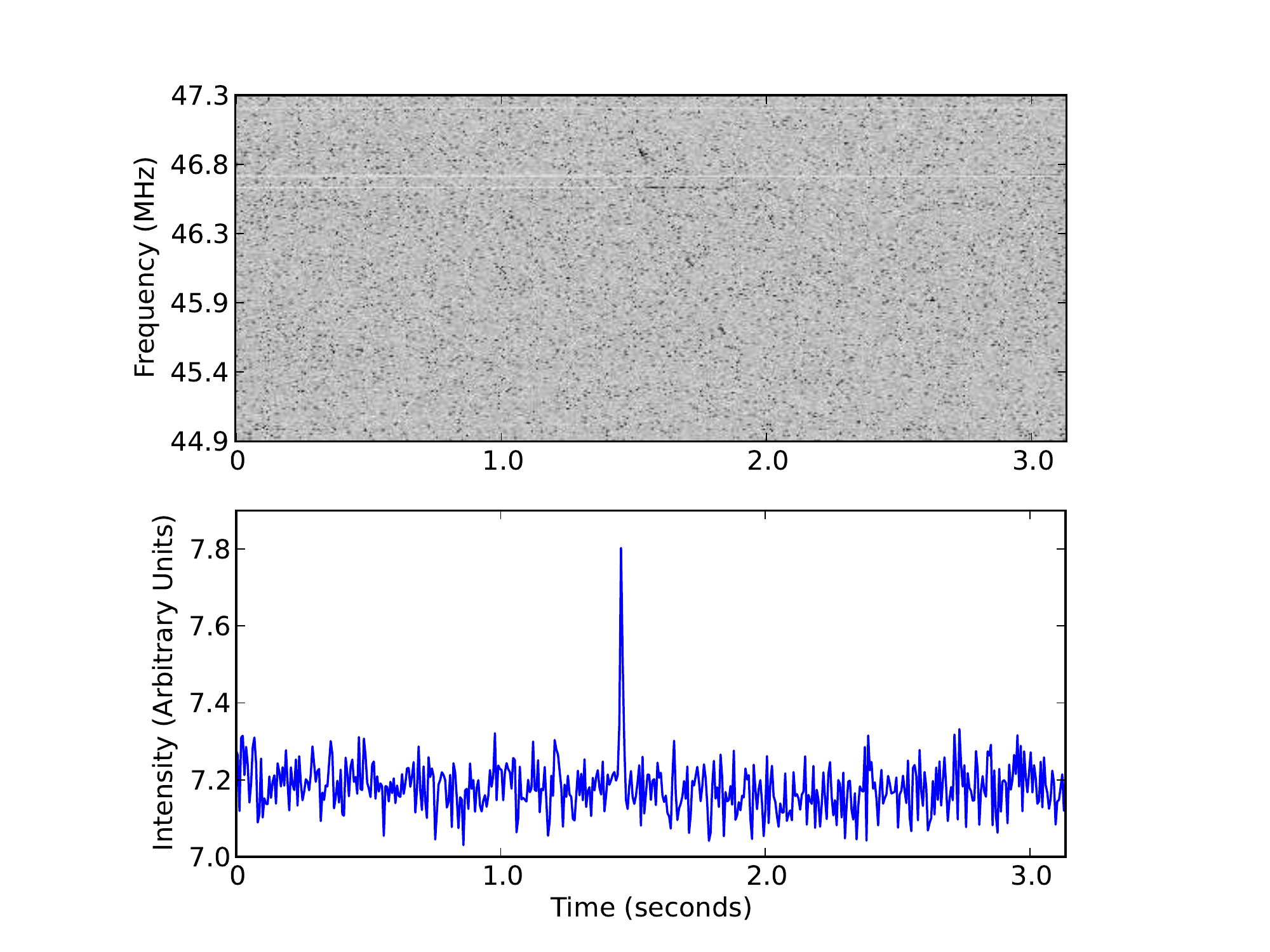}
\caption{A portion of a full spectrogram chosen to show a detected giant pulse, accompanied by the dedispersed time series constructed from the data displayed, for DM=2.97~pc~cm$^{-3}$.  The pulse can be seen in the displayed spectrogram as a series of narrow linear features running along a sloping, slight curved path from earlier arrival times at high frequency (starting at $\sim$1.4s) to later arrival times at low frequency (ending at $\sim$2s). The dedispersed time series is constructed by summing the displayed spectrogram values in frequency, after shifting each frequency in time to compensate for the later arrival time of lower frequencies.   Horizontal features can be seen in the spectrogram; these features are frequency bins masked due to RFI contamination. The SNR of the pulse in the displayed time series is about 8.8 (treating the pulse as an isolated event).  In the time series constructed from the full 16-MHz spectrogram, this pulse has SNR=21.7. The ratio of this pulse's SNR to that of the mean pulse is 16.4.}
\label{signalintimeseriesandspectrum}
\end{center}
\end{figure}

\begin{figure}
\begin{center}
\includegraphics[width=.75\textwidth]{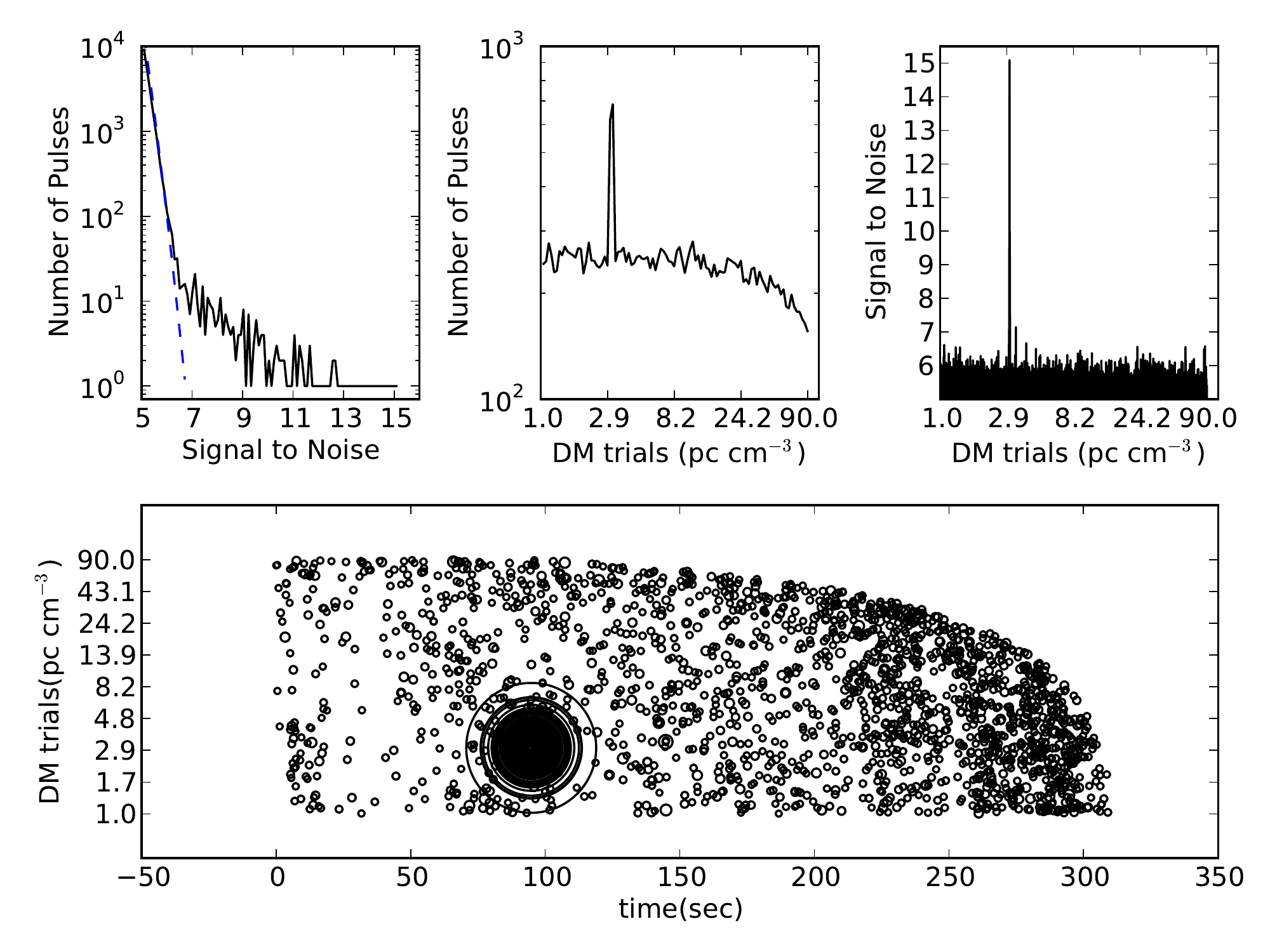}
\caption{A representative result from a large DM-space search for transient pulses in one of the 5-minute subsets of data, following the method of \citep{2003ApJ...596.1142C}.  Only events (generically called ``pulses'' by \citep{2003ApJ...596.1142C}) in the time series with SNR$>$5 are displayed. The upper left panel shows that the number of pulses versus SNR deviates from the result expected in the case of Gaussian noise (the dashed blue line), only for SNR$>$6.5. Thus events with SNR$<6.5$ cannot be distinguished from Gaussian noise in the time series, and should not be construed as actual pulses from PSR~B0950+08. A single giant pulse in this 5-minute observation will produce multiple events in the search across DMs similar to the pulsar's DM. The upper middle and upper right panels show a peak appearing at the appropriate DM for PSR~B0950+08. The upper middle panel displays the number of events found (with SNR$>5$) versus DM. The upper right panel is a plot of the SNR of each event found (with SNR$>5$) versus DM.  The lower panel shows a strong pulse appearing (as the large blackened circle) in the dedispersed time series for the appropriate DM; circle size is proportional to SNR. The absence of circles in the upper right of that lower panel is due to the time shifting inherent in the dedispersion procedure, applied to a data set of finite temporal duration. Due to limited resolution, not all pulses at the DM of PSR B0950+08 with SNR$>6.5$ can be seen in the lower panel. }
\label{GpsEventRate}
\end{center}
\end{figure}

\begin{figure}
\begin{center}
\includegraphics[width=.75\textwidth]{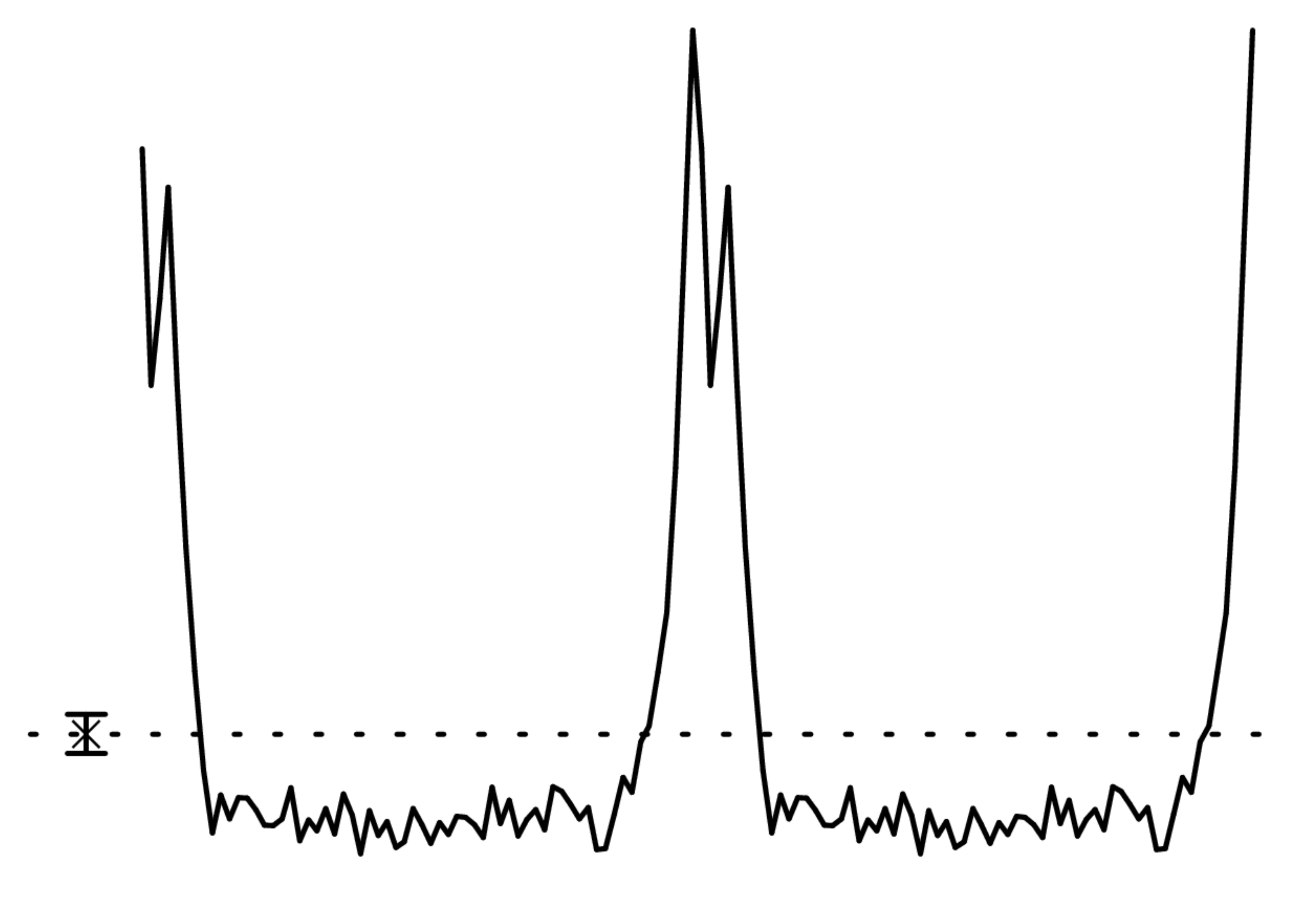}
\caption{The mean pulse profile obtained from PRESTO \citep{RansomThesis}, for one of the 6-hour observing sessions.  Note that the time span displayed is two pulse periods. The dashed line is the average intensity in the profile, the error bars represent the rms of all the intensity values.}
\label{PRESTOprofile}
\end{center}
\end{figure}

\begin{figure}
\begin{center}
\includegraphics[width=.75\textwidth]{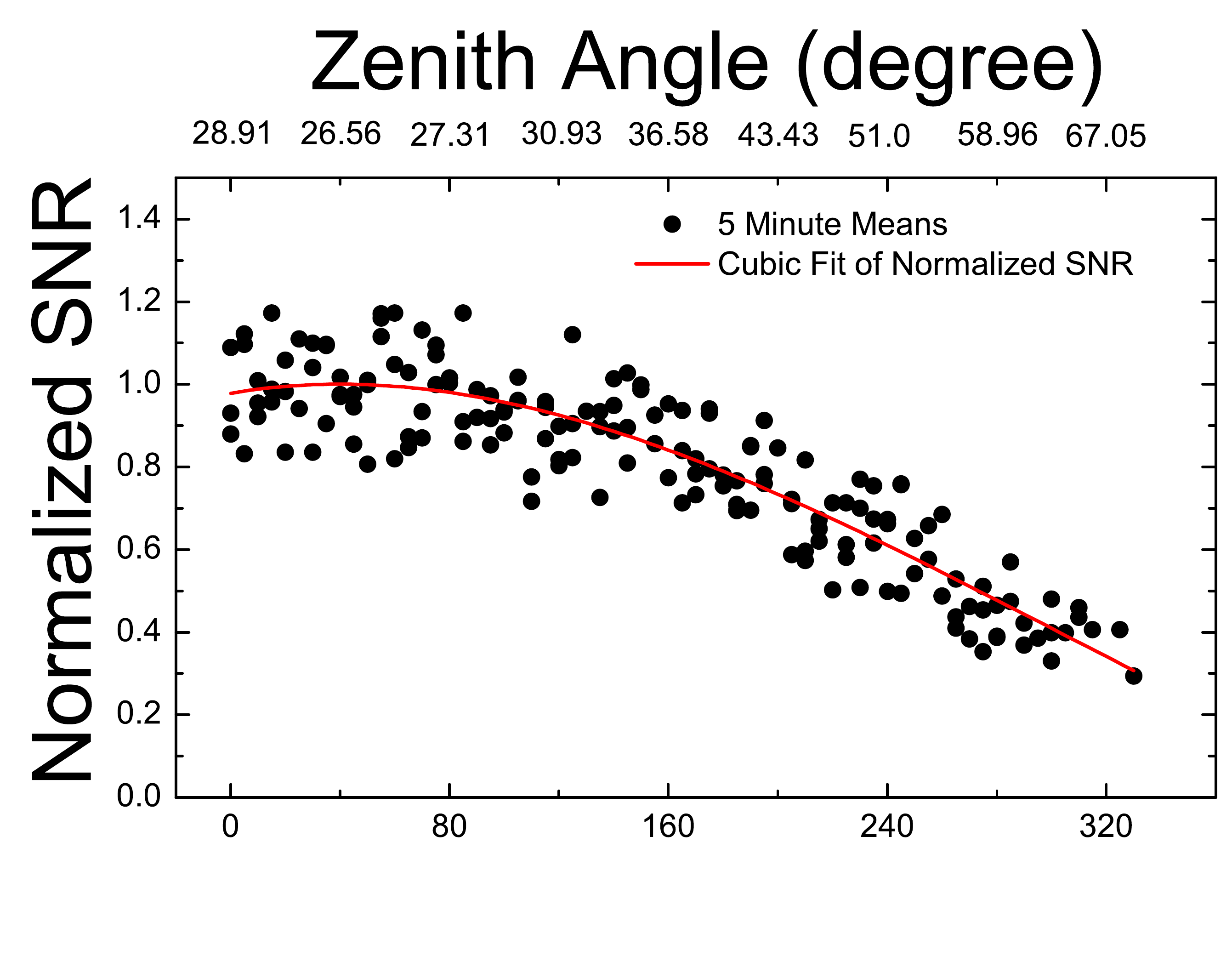}
\caption{The average SNR in 5-minute intervals, during three of the 6-hour observing sessions.  A few 5-minute averages were removed since they deviated significantly from the typical result, due to the presence of especially strong GPs.  A third-order polynomial fit to the data is shown.  The SNRs have been normalized, such that the fit peaks at a value of 1, which occurred at about the transit time of PSR~B0950+08. The PSR~B0950+08 pass meridian with zenith angle = 26.15 or 48 minutes after first frame of observation.}
\label{meanpulsevstime}
\end{center}
\end{figure}

\begin{figure}
\begin{center}
\includegraphics[width=.75\textwidth]{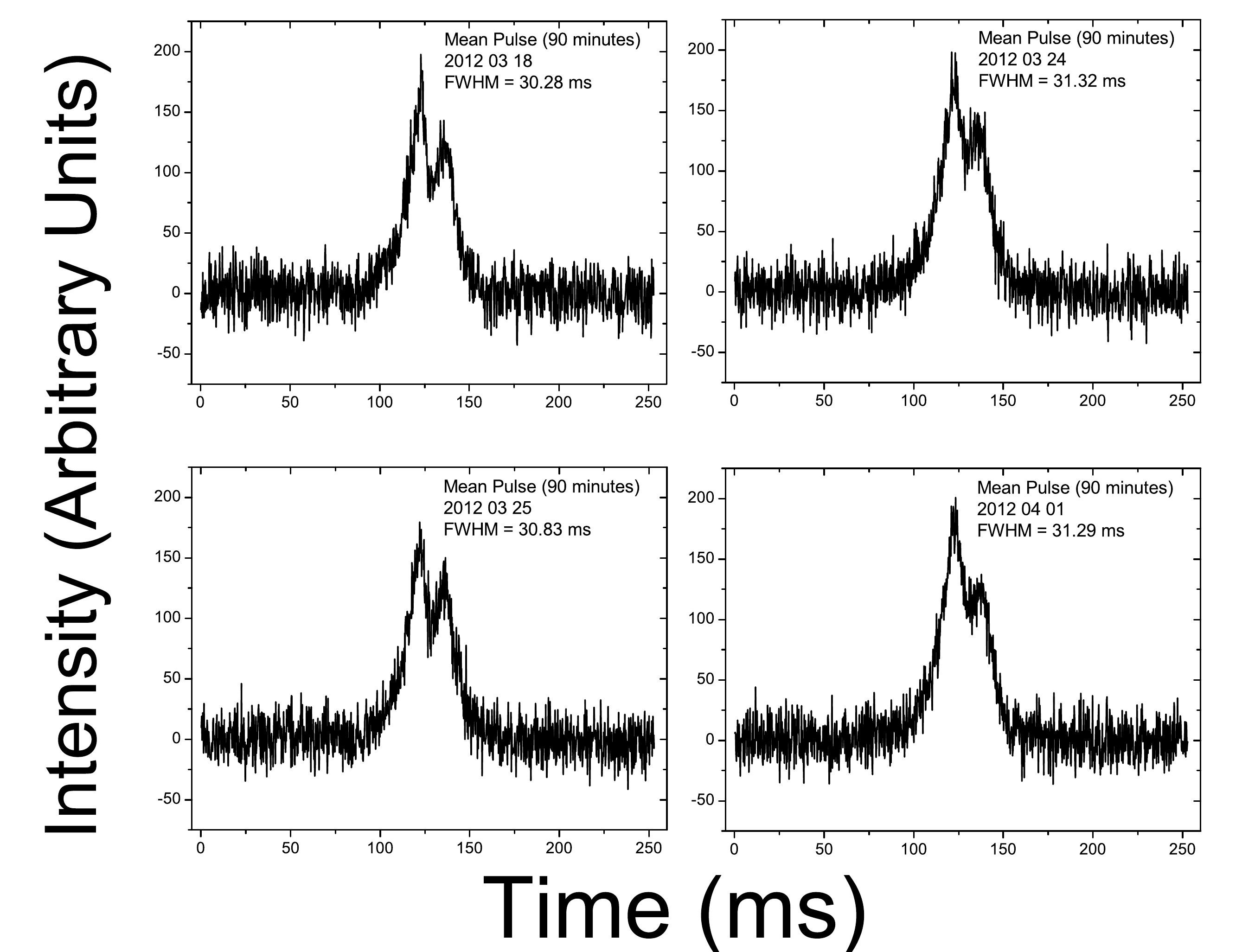}
\caption{Mean pulse profiles from the first 90 minutes of each of the four observing sessions, when PSR~B0950+08 was near transit. The time span displayed is one pulse period. The average FWHM pulse width is approximately 30.5~ms, as determined by fitting a single Gaussian to each of these profiles.}
\label{meanpulses}
\end{center}
\end{figure}

\begin{figure}
\begin{center}
\includegraphics[width=0.75\textwidth]{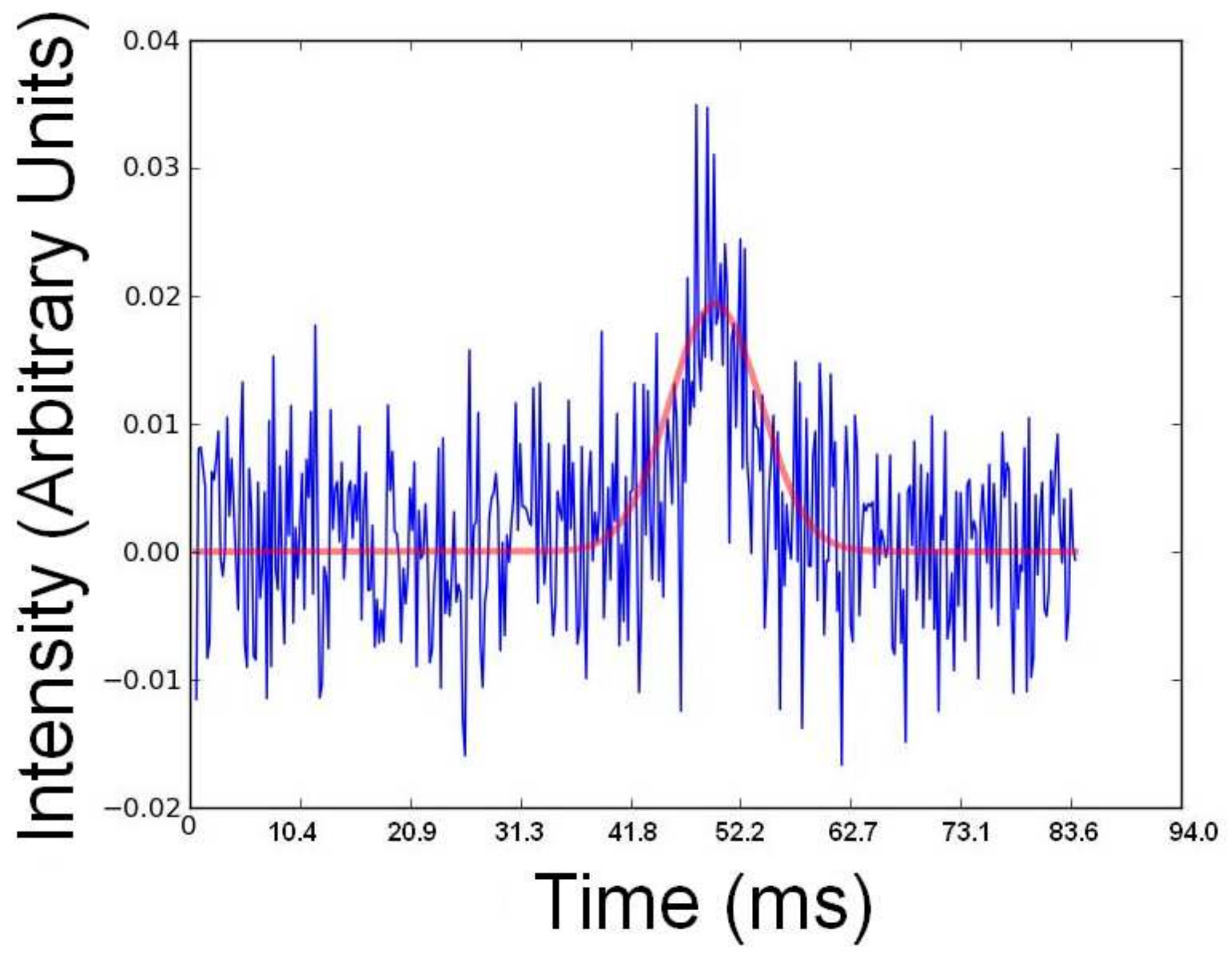}
\caption{A GP in a dedispersed time series.  Also shown is a Gaussian fit to the pulse.}
\label{gpidentifygaussianfit}
\end{center}
\end{figure}

\begin{figure}
\begin{center}
\includegraphics[width=0.75\textwidth]{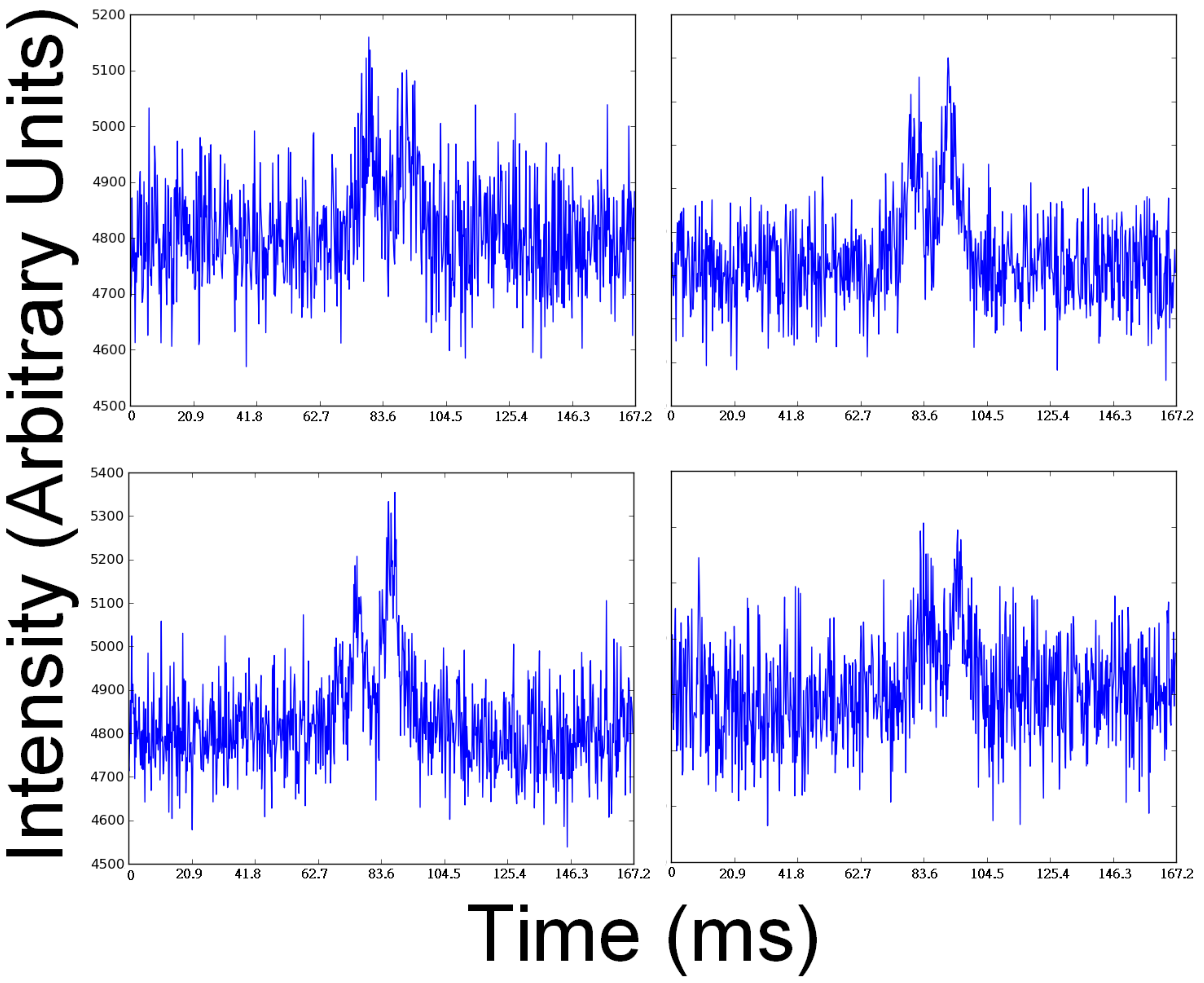}
\caption{Examples of GPs with double peaks in the dedispersed time series. The typical time difference between peaks is about 10 ms. }
\label{Gps2peaks}
\end{center}
\end{figure}

\begin{figure}
\begin{center}
\includegraphics[width=0.75\textwidth]{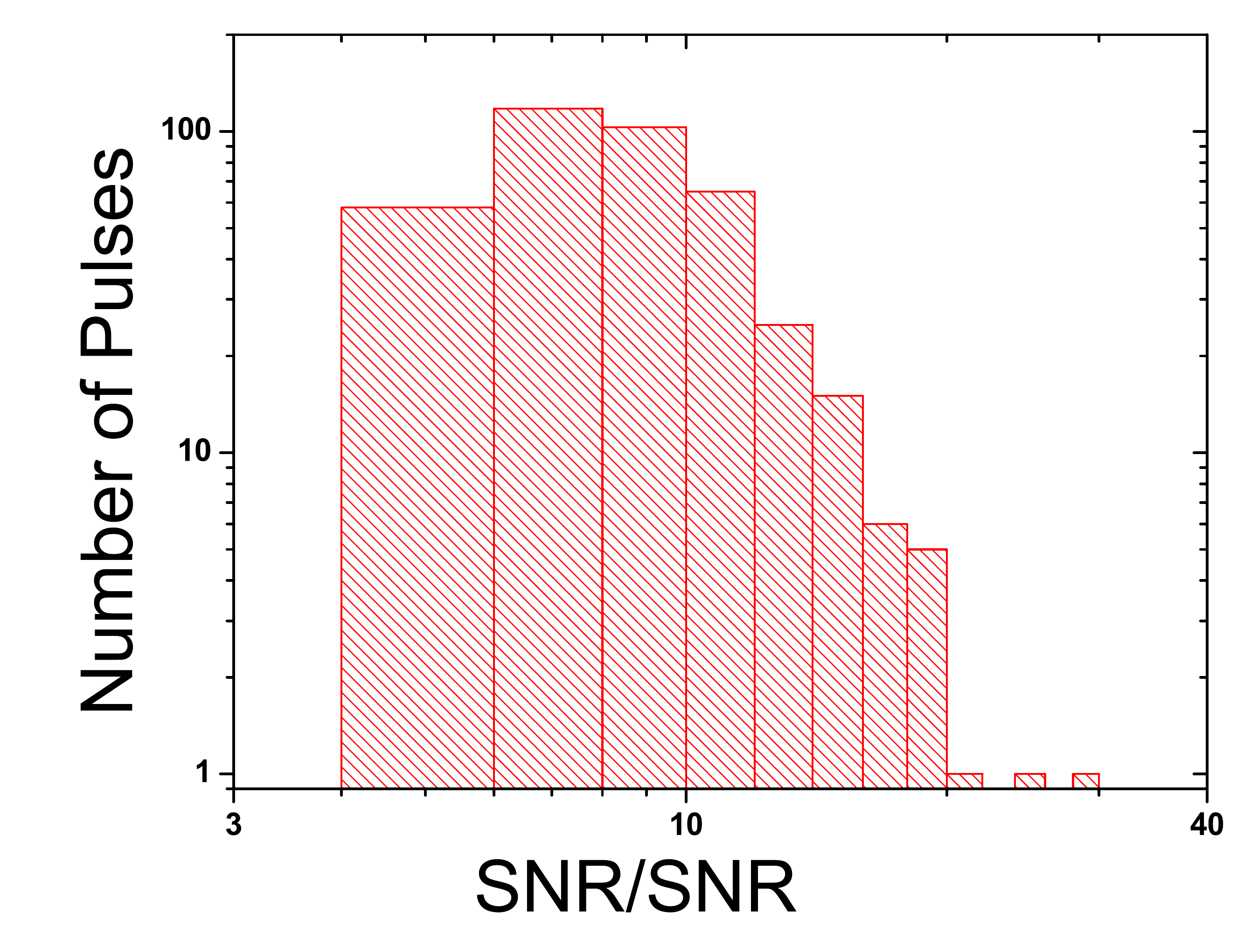}
\caption{The histogram of detected pulses, given in terms of pulse SNR relative to that of the mean pulse.  Giant pulses are those that have SNRs that exceed the mean pulse SNR by at least a factor of 10.}
\label{pulsenumberhistogram}
\end{center}
\end{figure}


\begin{figure}
\begin{center}
\includegraphics[width=0.75\textwidth]{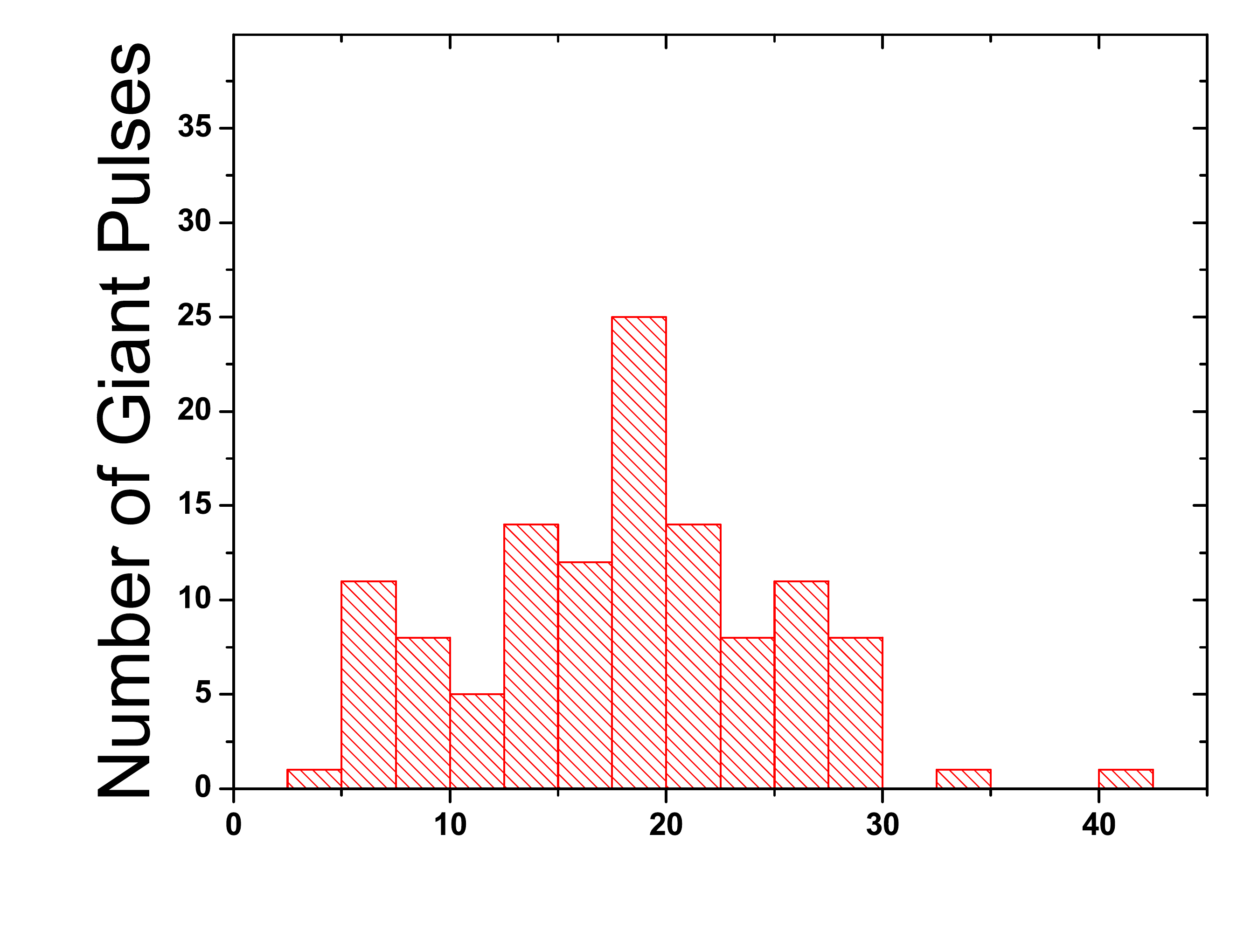}
\caption{The histogram of temporal widths (FWHM) for single-Gaussian fits to the individual giant pulses.  The average value is 17.8~ms.  For comparison, the FWHM value for a Gaussian fit to the mean pulse is 30.5~ms.}
\label{widthhistogram}
\end{center}
\end{figure}

\begin{figure}
\begin{center}
\includegraphics[width=0.75\textwidth]{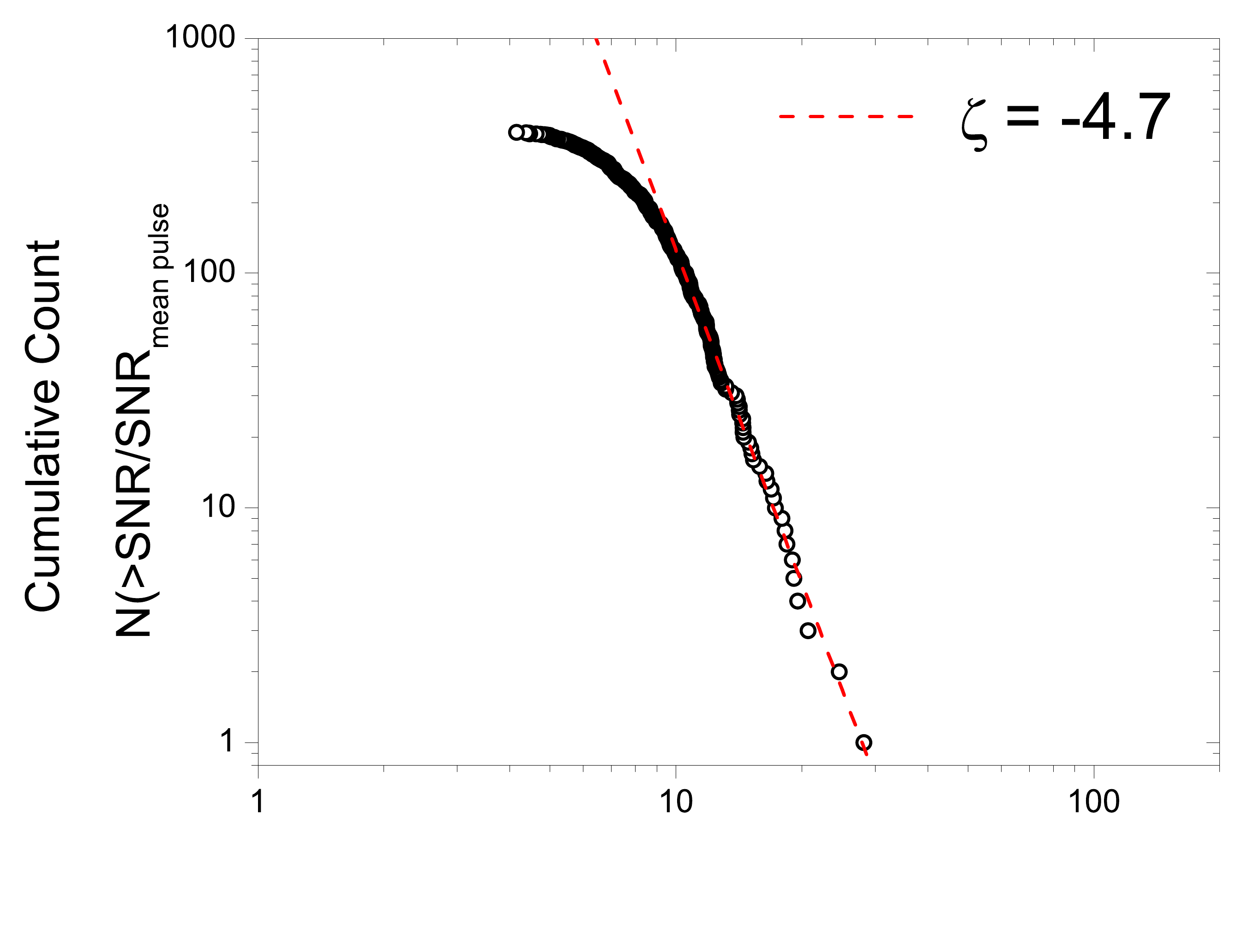}
\caption{The cumulative number of pulse strength $N(>S)$ where $S$ is the SNR relative to the mean pulse SNR, i.e., $S= SNR/SNR_{\rm mean\ pulse}$. The number of pulses per hour with relative SNR exceeding the value $S$ can be determined by dividing $N(>S)$ by 24~hours.  A power-law fit to GPs only ($S >10$), yields $N(>S)\propto S^\zeta$, where $\zeta=-4.7$.
}
\label{loglogGpsEventRate}
\end{center}
\end{figure}


\begin{figure}
\begin{center}
\includegraphics[width=0.75\textwidth]{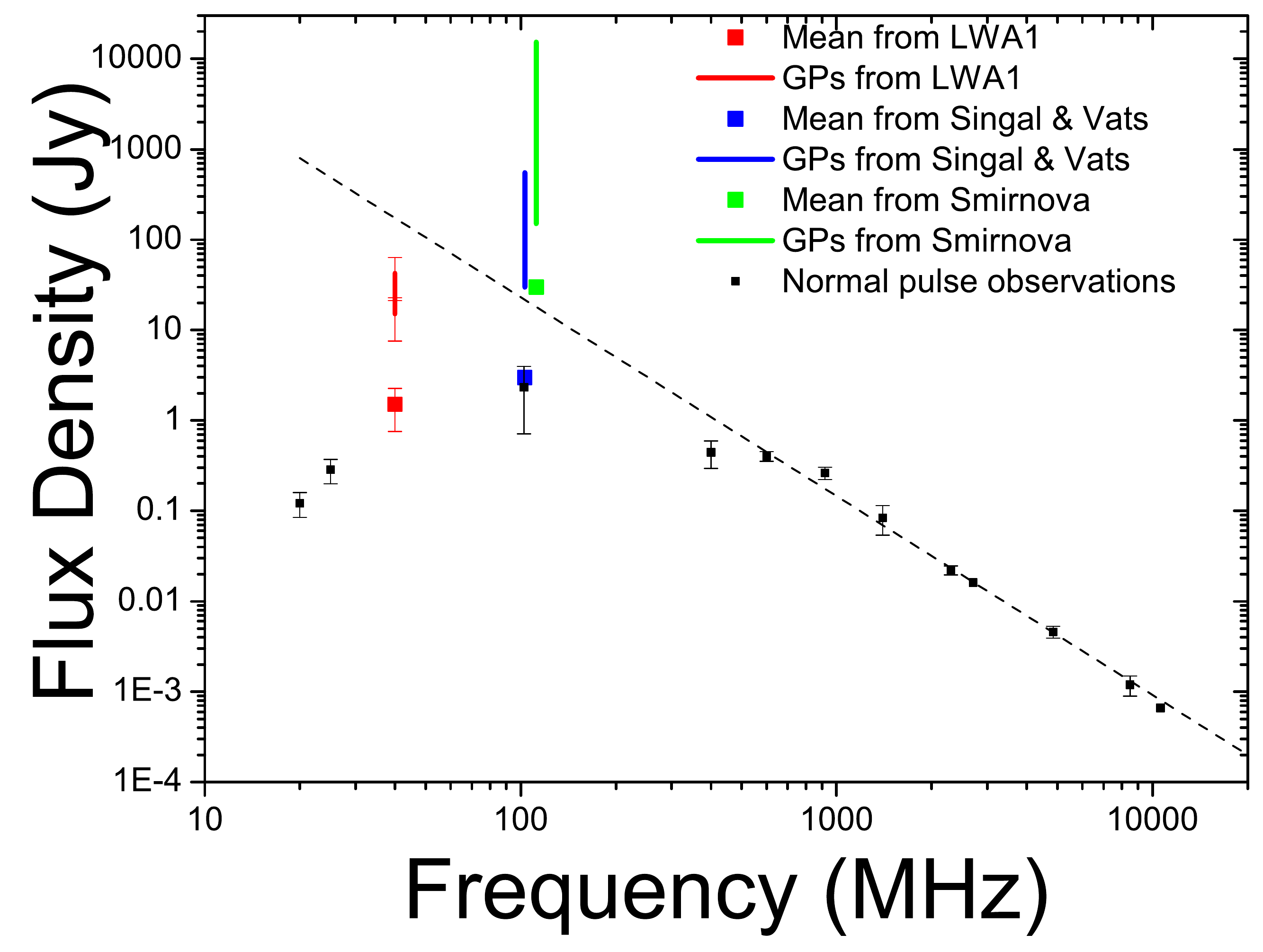}
\caption{Observed flux densities for PSR~B0950+08. Red symbols are for LWA1 at 39.4~MHz (this work); the red square is for the mean pulse, with 50\% error bars.  The thick red vertical line indicates the range of GP values; the thin error bars extending above and below indicate the 50\% error associated with the lowest GP and highest GP flux density. Blue and green symbols similarly represent observations by \citep{2012AJ....144..155S} at 103~MHz, and by \citet{2012ARep...56..430S} at 112~MHz. The normal pulse observations (black squares) are from (left to right: points 1--3, \citep{2013MNRAS.431.3624Z}; points 4--6, and 9, \citep{GouldLyne}; point 7, 10, and 12, \citep{Seiradakis}; points 8 and 11, \citep{vonHoensbroechXilouris}.  The dashed black line is a fit through the normal pulse observations, of spectral index $-$2.2.
} 
\label{SpectralIndex}
\end{center}
\end{figure}

\begin{figure}
\begin{center}
\includegraphics[width=0.75\textwidth]{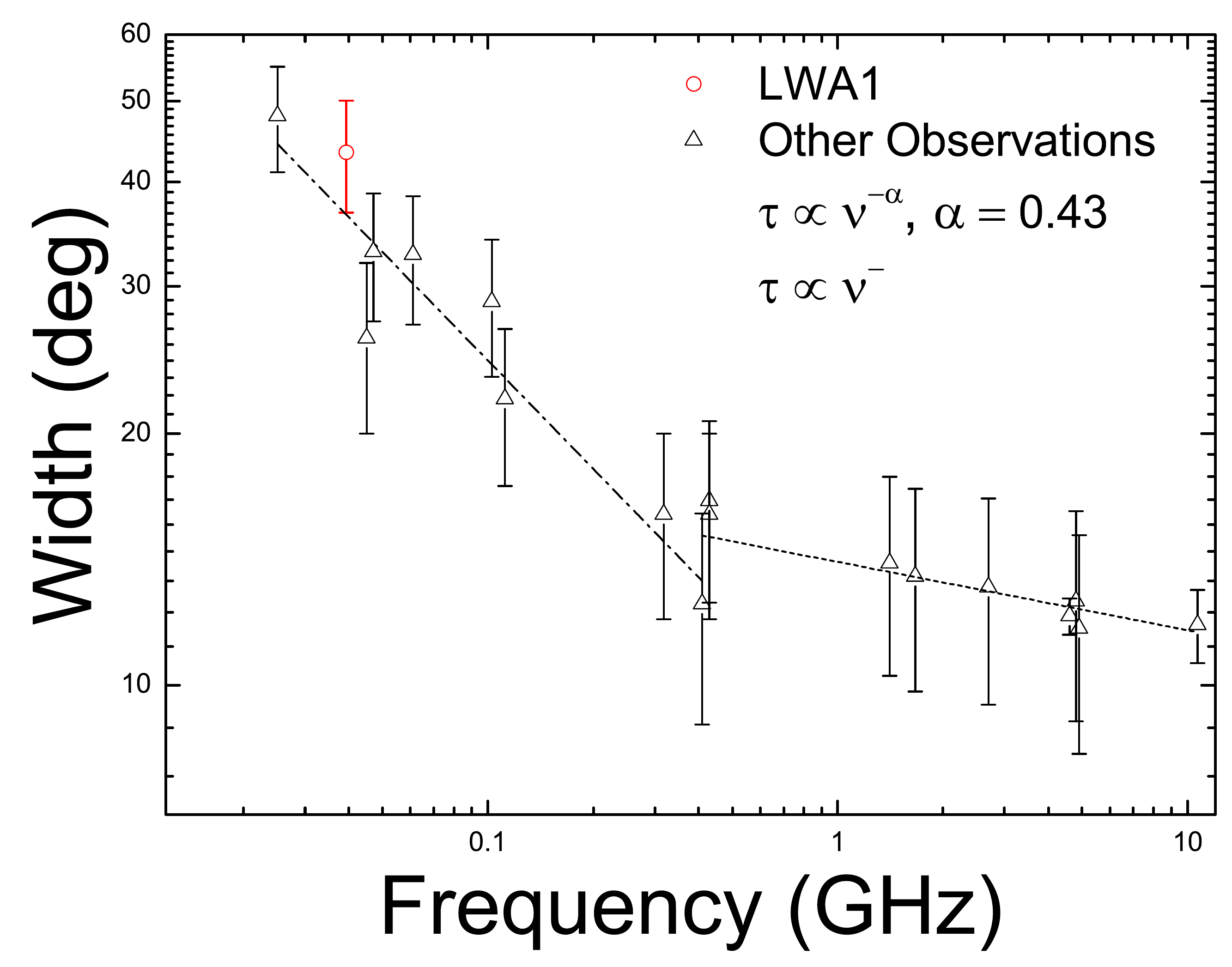}
\caption{The spectra of pulse widths observed for PSR~B0950+08 normalized to have a period of $360^{\circ}$ with the current observation at 39.4MHz included. We fit the spectral index for two frequency ranges, above and below 430MHz. The vertical error bar is one $\sigma$ assuming Poisson statistics unless otherwise specified in references for the observations cited. \citet{1971ApJS...23..283M}:0.41GHz, 1.665GHz. \citep{1975A&A....38..169S}:2.7GHz, 4.9GH. 
\citet{1979SvA....23..179I}: 61MHz, 102.5MHz.
\citet{1981AJ.....86..418R}: 430MHz.
\citet{1986A&A...161..183K}: 4.6GHz, 10.7GHz.
\citet{1992ApJ...385..273P}: 25MHz, 47MHz, 112MHz, 430MHz, 1408MHz, 4800MHz.
\citet{1995A&A...301..182R}: 45MHz.
This work: 39.4MHz.
}  
\label{pulsewidthindex}
\end{center}
\end{figure}

\end{document}